\documentclass[11pt]{article}
\usepackage{graphicx}
\usepackage{amsmath}
\usepackage{amsfonts}
\usepackage{amssymb}
\newtheorem{theorem}{Theorem}

\newtheorem{corollary}[theorem]{Corollary}

\newtheorem{definition}[theorem]{Definition}

\newtheorem{proposition}[theorem]{Proposition}

\newenvironment{proof}[1][Proof]{\textbf{#1.} }{\ \rule{0.5em}{0.5em}}

\begin{document}

\title{A Laguerre Polynomial Orthogonality and the Hydrogen Atom}
\author{Charles F. Dunkl\thanks{During the preparation of this paper the author was
partially supported by NSF grants DMS 9970389 and DMS 0100539.}\\Department of Mathematics\\University of Virginia}
\date{May 17, 2002}
\maketitle
\begin{abstract}
The radial part of the wave function\ of an electron in a Coulomb potential is
the product of a Laguerre polynomial and an exponential with the variable
scaled by a factor depending on the degree. This note presents an elementary
proof of the orthogonality of wave functions with differing energy levels. It
is also shown that this is the only other natural orthogonality for Laguerre
polynomials. By expanding in terms of the usual Laguerre polynomial basis an
analogous strange orthogonality is obtained for Meixner polynomials.
\end{abstract}

\section{Introduction}

The time-independent Schr\"{o}dinger equation for one electron subject to a
Coulomb potential is
\[
-\frac{\hbar^{2}}{2m}\Delta\psi\left(  x\right)  -\frac{c}{||x||}\psi\left(
x\right)  =E\psi\left(  x\right)  ,
\]
for certain constants $m,c$ (see Robinett \cite{Rob}, Ch. 18). Rewrite the
equation as
\[
\Delta\psi\left(  x\right)  +\frac{k}{||x||}\psi\left(  x\right)
=-\lambda\psi\left(  x\right)  ,
\]
where $k$ is a positive parameter and $\lambda$ denotes the rescaled energy
level; let $x\in\mathbb{R}^{N}$ (usually $N=3$, of course) and $||x||=\left(
\sum_{j=1}^{N}x_{j}^{2}\right)  ^{1/2}$. We separate variables by considering
solutions of the form $\psi\left(  x\right)  =\phi\left(  ||x||\right)
p_{l}\left(  x\right)  $ where $p_{l}$ is a harmonic homogeneous polynomial of
degree $l=0,1,2,\ldots$ (the angular momentum quantum number). Then $\phi$
must satisfy the equation
\begin{equation}
\frac{d^{2}}{dr^{2}}\phi\left(  r\right)  +\frac{N+2l-1}{r}\frac{d}{dr}%
\phi\left(  r\right)  +\left(  \frac{k}{r}+\lambda\right)  \phi\left(
r\right)  =0.\label{radeq}%
\end{equation}
Using spherical polar coordinates for $\mathbb{R}^{N}$ we see that $\phi$ must
be square-integrable for the measure $r^{2l+N-1}dr$ on $\{r:r>0\}$ (we
restrict our attention to the point spectrum). Furthermore solutions for the
same value of $l$ but different energy levels must be orthogonal for this
measure, by the self-adjointness of the equation. The topic of this note is a
direct elementary proof of this orthogonality. It will also be shown that
these functions and the standard set form the only possible (in a sense to be
specified) orthogonal sets of scaled Laguerre polynomials. Then by expanding
the functions in terms of the usual Laguerre basis an analogous orthogonality
is obtained for Meixner polynomials. This orthogonal set incorporates a
parameter dependent on the degree and is a set of eigenfunctions of a
self-adjoint second-order difference operator resembling equation (\ref{radeq}).

In equation (\ref{radeq}) set $\lambda=-\mu^{2}$ (the bound energy levels are
negative) and substitute $\phi\left(  r\right)  =e^{-\mu r}g\left(  2\mu
r\right)  $, then $g\left(  s\right)  $ satisfies
\[
\frac{d^{2}}{ds^{2}}g\left(  s\right)  +\left(  N+2l-1-s\right)  \frac{d}%
{ds}g\left(  s\right)  +\frac{1}{2}\left(  1+\frac{k}{\mu}-N-2l\right)
g\left(  s\right)  =0.
\]
The integrable solutions are obtained by setting $\frac{1}{2}\left(
1+\frac{k}{\mu}-N-2l\right)  =n$ for some $n=0,1,2,3,\ldots$ then $\phi\left(
r\right)  =L_{n}^{\alpha}\left(  2\mu_{n}r\right)  \exp\left(  -\mu
_{n}r\right)  $ where $\mu_{n}=\frac{k}{2\left(  \alpha+2n+1\right)  }$ and
$\alpha=N+2l-2$ ; see (\ref{lageq}) below ($\phi$ is not normalized). The
energy levels are $-\mu_{n}^{2}$ (specifically $-\left(  \frac{k}{4\left(
n+l+1\right)  }\right)  ^{2}$ for $N=3$).

These radial Laguerre-type functions continue to be useful in the research
literature: Beckers and Debergh \cite{BD} used them for supersymmetric
properties of the parastatistical hydrogen atom, Blaive and Cadilhac \cite{BC}
derived an expansion in powers of $\mu_{n}$ (for fixed $l$), Dehesa et al
\cite{DY} found approximations for entropies which involve integrals of the
logarithms of the Laguerre polynomials, Xu and Kong \cite{XK} constructed
shift operators for the radial Hamiltonian.

The Pochhammer symbol is defined by
\[
\left(  a\right)  _{n}=\prod_{i=1}^{n}\left(  a+i-1\right)  ,\,a\in\mathbb{R},
\]
and the hypergeometric ($\,_{2}F_{1}$) series is
\[
_{2}F_{1}\left(  a,b;c;x\right)  =\sum_{j=0}^{\infty}\frac{\left(  a\right)
_{j}\left(  b\right)  _{j}}{\left(  c\right)  _{j}\,j!}x^{j}.
\]

The basic facts about the Laguerre polynomial of index $\alpha$ and degree $n
$ that will be needed are:
\begin{gather}
L_{n}^{\alpha}\left(  x\right)  =\frac{\left(  \alpha+1\right)  _{n}}{n!}%
\sum_{j=0}^{n}\frac{\left(  -n\right)  _{j}}{\left(  \alpha+1\right)  _{j}%
}\frac{x^{j}}{j!},\nonumber\\
\int_{0}^{\infty}L_{m}^{\alpha}\left(  x\right)  L_{n}^{\alpha}\left(
x\right)  x^{\alpha}e^{-x}dx=\delta_{mn}\Gamma\left(  \alpha+1\right)
\frac{\left(  \alpha+1\right)  _{n}}{n!},\label{ortho}\\
\left(  x\frac{d^{2}}{dx^{2}}+(\alpha+1-x)\frac{d}{dx}+n\right)  L_{n}%
^{\alpha}\left(  x\right)  =0,\label{lageq}\\
\left(  n+1\right)  L_{n+1}^{\alpha}\left(  x\right)  -\left(  \alpha
+1+2n-x\right)  L_{n}^{\alpha}\left(  x\right)  +\left(  n+\alpha\right)
L_{n-1}^{\alpha}\left(  x\right)  =0.\label{lagrec}%
\end{gather}
Their proofs and further information are in Szeg\"{o} \cite{Sze}, Ch.5. The
author gratefully acknowledges a useful discussion with R. Y\'{a}\~{n}ez.

\section{Orthogonality relation}

Fix a parameter $\alpha>-1.$ The scaling factor $k$ appearing in equation
(\ref{radeq}) is set equal to 1.

\begin{definition}
\label{defphi}For $n=0,1,2,\ldots$and $x\in\mathbb{R}$ let
\[
\phi_{n}\left(  x\right)  =L_{n}^{\alpha}\left(  \frac{x}{\alpha+2n+1}\right)
\exp\left(  -\frac{x}{2\left(  \alpha+2n+1\right)  }\right)  .
\]
\end{definition}

\begin{theorem}
\label{normphi}For $m,n=0,1,2,\ldots$ the following orthogonality relation
holds:
\[
\int_{0}^{\infty}\phi_{m}\left(  x\right)  \phi_{n}\left(  x\right)
x^{\alpha+1}\,dx=\frac{\delta_{mn}}{n!}\left(  \alpha+2n+1\right)  ^{\alpha
+3}\Gamma\left(  \alpha+1+n\right)  .
\]
\end{theorem}

The proof is composed of two parts. First, the case $m=n$ follows from the
orthogonality relation (\ref{ortho}) and the three-term recurrence
(\ref{lagrec}) in the form
\[
xL_{n}^{\alpha}\left(  x\right)  =-\left(  n+1\right)  L_{n+1}^{\alpha}\left(
x\right)  +\left(  \alpha+2n+1\right)  L_{n}^{\alpha}\left(  x\right)
+\left(  \alpha+n\right)  L_{n-1}^{\alpha}\left(  x\right)  ,
\]
and thus $\int_{0}^{\infty}xL_{n}^{\alpha}\left(  x\right)  ^{2}x^{\alpha
}e^{-x}\,dx=\left(  \alpha+2n+1\right)  \Gamma\left(  \alpha+1+n\right)  /n!$.
In the integral $\int_{0}^{\infty}\phi_{n}\left(  x\right)  ^{2}x^{\alpha
+1}\,dx$ change the variable to $s=\frac{x}{\alpha+2n+1}$.

Second, fix $m\neq n$ and in the integral $\int_{0}^{\infty}\phi_{m}\left(
x\right)  \phi_{n}\left(  x\right)  x^{\alpha+1}\,dx$ change the variable to
\[
s=\frac{x}{2}\left(  \frac{1}{\alpha+2m+1}+\frac{1}{\alpha+2n+1}\right)  ,
\]
and let
\[
\beta=\frac{\alpha+2n+1}{\alpha+m+n+1};
\]
then the desired integral equals $\left(  \dfrac{\left(  \alpha+2m+1\right)
\left(  \alpha+2n+1\right)  }{\alpha+m+n+1}\right)  ^{\alpha+2}$ times
\begin{equation}
\int_{0}^{\infty}L_{m}^{\alpha}\left(  \beta s\right)  L_{n}^{\alpha}\left(
\left(  2-\beta\right)  s\right)  s^{\alpha+1}e^{-s}ds.\label{zprod}%
\end{equation}
We will show this equals zero by expressing the two polynomials in terms of
$L_{j}^{\alpha+1}\left(  s\right)  $, which simplifies evaluation of the
integral. Recall the identities
\begin{align}
L_{n}^{\alpha}\left(  x\right)   &  =L_{n}^{\alpha+1}\left(  x\right)
-L_{n-1}^{\alpha+1}\left(  x\right)  ,\label{contig}\\
L_{n}^{\delta}\left(  bx\right)   &  =\sum_{j=0}^{n}\frac{\left(
\delta+1+j\right)  _{n-j}}{\left(  n-j\right)  !}b^{j}\left(  1-b\right)
^{n-j}L_{j}^{\delta}\left(  x\right)  ,\label{rescale}%
\end{align}
which hold for arbitrary $x,b\in\mathbb{R}$, $\delta>-1$ and $n=0,1,2,\ldots$.
Then (with $\delta=\alpha+1$ in (\ref{rescale}))
\begin{align*}
L_{m}^{\alpha}\left(  \beta s\right)   &  =L_{m}^{\alpha+1}\left(  \beta
s\right)  -L_{m-1}^{\alpha+1}\left(  \beta s\right) \\
&  =\sum_{j=0}^{m}\frac{\left(  \alpha+2+j\right)  _{m-1-j}}{\left(
m-j\right)  !}\beta^{j}\left(  1-\beta\right)  ^{m-1-j}L_{j}^{\alpha+1}\left(
s\right) \\
&  \times\left(  \left(  1-\beta\right)  \left(  \alpha+1+m\right)  -\left(
m-j\right)  \right) \\
&  =\sum_{j=0}^{m}\frac{\left(  \alpha+2+j\right)  _{m-1-j}}{\left(
m-j\right)  !}\beta^{j}\left(  1-\beta\right)  ^{m-1-j}\left(  j-n\left(
2-\beta\right)  \right)  L_{j}^{\alpha+1}\left(  s\right)
\end{align*}
and
\begin{align*}
L_{n}^{\alpha}\left(  \left(  2-\beta\right)  s\right)   &  =L_{n}^{\alpha
+1}\left(  \left(  2-\beta\right)  s\right)  -L_{n-1}^{\alpha+1}\left(
\left(  2-\beta\right)  s\right) \\
&  =\sum_{j=0}^{n}\frac{\left(  \alpha+2+j\right)  _{n-1-j}}{\left(
n-j\right)  !}\left(  2-\beta\right)  ^{j}\left(  \beta-1\right)
^{n-1-j}L_{j}^{\alpha+1}\left(  s\right) \\
&  \times\left(  \left(  \beta-1\right)  \left(  \alpha+1+n\right)  -\left(
n-j\right)  \right) \\
&  =\sum_{j=0}^{n}\frac{\left(  \alpha+2+j\right)  _{n-1-j}}{\left(
n-j\right)  !}\left(  2-\beta\right)  ^{j}\left(  \beta-1\right)
^{n-1-j}\left(  j-m\beta\right)  L_{j}^{\alpha+1}\left(  s\right)  .
\end{align*}
Now multiply the two expressions together and integrate with respect to the
measure $s^{\alpha+1}e^{-s}ds$ on $\{s:s>0\}$. By the orthogonality
(\ref{ortho}) with parameter $\alpha+1$ integral (\ref{zprod}) equals
$\Gamma\left(  \alpha+2\right)  $ times
\begin{align}
&  \frac{\left(  \alpha+2\right)  _{m-1}\left(  \alpha+2\right)  _{n-1}%
}{m!\,n!}\left(  1-\beta\right)  ^{m-1}\left(  \beta-1\right)  ^{n-1}%
\label{bsum}\\
&  \times\sum_{j=0}^{\min\left(  m,n\right)  }\frac{\left(  -m\right)
_{j}\left(  -n\right)  _{j}}{\left(  \alpha+2\right)  _{j}\,j!}\left(
\frac{\beta\left(  \beta-2\right)  }{\left(  1-\beta\right)  ^{2}}\right)
^{j}\left(  j-n\left(  2-\beta\right)  \right)  \left(  j-m\beta\right)
.\nonumber
\end{align}
The formula holds when one of $m,n=0$ by the convention $\left(
\alpha+2\right)  _{-1}=\frac{1}{\alpha+1}.$ Let
\[
\gamma=\frac{\beta\left(  \beta-2\right)  }{\left(  1-\beta\right)  ^{2}%
}=1-\frac{1}{\left(  1-\beta\right)  ^{2}}%
\]
then
\[
\left(  j-n\left(  2-\beta\right)  \right)  \left(  j-m\beta\right)  =j\left(
\alpha+1+j\right)  +\frac{\gamma}{\gamma-1}\left(  mn-j\left(  \alpha
+m+n+1\right)  \right)  ,
\]
since $\beta\left(  2-\beta\right)  =\frac{\gamma}{\gamma-1}$. The second line
of the sum (\ref{bsum}) can be written as
\begin{align*}
&  mn\gamma\,_{2}F_{1}\left(  1-m,1-n;\alpha+2;\gamma\right)  +\frac{\gamma
mn}{\gamma-1}\,_{2}F_{1}\left(  -m,-n;\alpha+2;\gamma\right) \\
&  -\frac{\gamma^{2}mn\left(  \alpha+m+n+1\right)  }{\left(  \gamma-1\right)
\left(  \alpha+2\right)  }\,_{2}F_{1}\left(  1-m,1-n;\alpha+3;\gamma\right)  ;
\end{align*}
which equals $mn\frac{\gamma}{\gamma-1}$ times
\begin{align*}
&  \left(  \gamma-1\right)  \,_{2}F_{1}\left(  1-m,1-n;\alpha+2;\gamma\right)
+\,_{2}F_{1}\left(  -m,-n;\alpha+2;\gamma\right) \\
&  -\frac{\gamma\left(  \alpha+m+n+1\right)  }{\alpha+2}\,_{2}F_{1}\left(
1-m,1-n;\alpha+3;\gamma\right)  ;
\end{align*}
and this expression is zero by the contiguity relations for the $_{2}F_{1}%
$-series (by the straightforward calculation of the coefficient of $\gamma
^{j}$ for each $j$). This completes the proof of the theorem.

\section{Uniqueness}

We show that the only possible orthogonal sets $\left\{  L_{n}^{\alpha}\left(
\gamma_{n}x\right)  \exp\left(  -\frac{1}{2}\gamma_{n}x\right)  \right\}
_{n=0}^{\infty}$ for the measure $x^{\alpha+\kappa}dx$ on $\left\{
x:x>0\right\}  $ are the two cases (1) $\kappa=0,\,\gamma_{n}=\gamma_{0}$ for
all $n$; (2) $\kappa=1,\,\gamma_{n}=\gamma_{0}\frac{\alpha+1}{\alpha+2n+1}$.
By rescaling $x$ we may assume $\gamma_{0}=1$. The choice of the basis
functions is motivated by their general utility and precise asymptotic
formulae. Let $\phi_{n}\left(  x\right)  =L_{n}^{\alpha}\left(  \gamma
_{n}x\right)  \exp\left(  -\frac{1}{2}\gamma_{n}x\right)  $ (with each
$\gamma_{n}>0$) and define the inner product
\[
\left\langle f,g\right\rangle =\Gamma\left(  \alpha+\kappa+1\right)  ^{-1}%
\int_{0}^{\infty}f\left(  x\right)  g\left(  x\right)  x^{\alpha+\kappa}dx,
\]
for suitable functions $f,g$. By using the same method as in Section 2
(formula (\ref{zprod})) the inner product $\left\langle \phi_{m},\phi
_{n}\right\rangle $ is expressed as a multiple of
\[
\int_{0}^{\infty}L_{m}^{\alpha}\left(  \beta s\right)  L_{n}^{\alpha}\left(
\left(  2-\beta\right)  s\right)  e^{-s}s^{\alpha+\kappa}ds,
\]
where $\beta=\frac{2\gamma_{m}}{\gamma_{m}+\gamma_{n}}$. The evaluations are
straightforward algebraic calculations and were carried out in the computer
algebra system Maple$^{TM}$, which was also used to find the polynomials
$q_{2},q_{3}$ and their resultant.

\subsection{The case $\kappa\neq0,1$}

We show that the equations $\left\langle \phi_{0},\phi_{1}\right\rangle
=\left\langle \phi_{0},\phi_{2}\right\rangle =\left\langle \phi_{1},\phi
_{2}\right\rangle =0$ are incompatible unless $\kappa=0$ or $1$. The equation
$\left\langle \phi_{0},\phi_{1}\right\rangle =0$ implies $\gamma_{1}%
=\frac{\alpha+1}{\alpha+2\kappa+1}$. The condition $\left\langle \phi_{0}%
,\phi_{2}\right\rangle =0$ is equivalent to a certain quadratic equation
$q_{2}\left(  \gamma_{2}\right)  =0$ (not displayed here). The discriminant of
$q_{2}$ equals $16\kappa\left(  \alpha+\kappa+1\right)  \left(  \alpha
+2\right)  $. Integrability requires $\alpha+\kappa+1>0$ hence there are real
solutions for $\gamma_{2}$ only if $\kappa\geq0$. The conditions $\left\langle
\phi_{1},\phi_{2}\right\rangle =0$ and $\gamma_{1}=\frac{\alpha+1}%
{\alpha+2\kappa+1}$ are equivalent to a certain cubic equation $q_{3}\left(
\gamma_{2}\right)  =0$. The resultant of $q_{2},q_{3}$ (a polynomial in the
coefficients which is zero if and only if there exists a simultaneous root) is
found to be the product of $\kappa\left(  \kappa-1\right)  $ by a polynomial
in $\left(  \alpha+1\right)  $ and $\kappa$ with all coefficients nonnegative.
This shows that $q_{2}\left(  \gamma_{2}\right)  =0=q_{3}\left(  \gamma
_{2}\right)  $ is impossible for $\kappa\geq0,\alpha>-1$ unless $\kappa=0$ or
$1$.

\subsection{The case $\kappa=0$ or $1$}

For the case $\kappa=0$ the condition $\left\langle \phi_{0},\phi
_{n}\right\rangle =0$ (for $n>0$) and the formula (\ref{rescale}) with
$\delta=\alpha$ show that $\left(  \beta-1\right)  ^{n}=0$ where $\beta
=\frac{2}{1+\gamma_{n}}$, hence $\gamma_{n}=1=\gamma_{0}$ for all $n$ (the
variable $\beta$ is not identical to the one used in Section 2, because the
scaling parameters $\gamma_{n}$ are initially undetermined).

For the case $\kappa=1$ , $\left\langle \phi_{0},\phi_{1}\right\rangle =0$
implies $\gamma_{1}=\frac{\alpha+1}{\alpha+3}$; further for $n\geq2$ the
condition $\left\langle \phi_{0},\phi_{n}\right\rangle =0$ and formulas
(\ref{contig}), (\ref{rescale}) imply
\[
\left(  \beta-1\right)  ^{n-1}\left(  \left(  \alpha+1+n\right)  \left(
\beta-1\right)  -n\right)  =0,
\]
(where $\beta=\frac{2}{1+\gamma_{n}}$) with the solutions $\beta=1$ and
$\frac{\alpha+2n+1}{\alpha+n+1}$. Thus $\gamma_{n}=\frac{\alpha+1}%
{\alpha+2n+1}$ or $1$. But $\gamma_{1}=\frac{\alpha+1}{\alpha+3}$, and the
computation of $\left\langle \phi_{1},\phi_{n}\right\rangle $ with $\gamma
_{n}=1$ yields
\[
\left\langle \phi_{1},\phi_{n}\right\rangle =\left(  -1\right)  ^{n-1}n\left(
\left(  \alpha+3\right)  n-1\right)  \frac{\left(  \alpha+3\right)  \left(
\alpha+3\right)  _{n-2}}{\left(  \alpha+1\right)  ^{n-2}\left(  \alpha
+2\right)  }%
\]
which is nonzero for $n\geq2$. This rules out the possibility that $\gamma
_{n}=1$ and shows that $\gamma_{n}=\frac{\alpha+1}{\alpha+2n+1}$ for all $n$
is necessary. The sufficiency was proved already.

\section{Change of Basis and Meixner Polynomials}

Since there is another obvious orthogonal basis for $L^{2}\left(  x^{\alpha
+1}dx\right)  $, the standard Laguerre functions $\left\{  L_{m}^{\alpha
+1}\left(  cx\right)  e^{-cx/2}:m=0,1,2,\ldots\right\}  $, with some scaling
parameter $c$, the computation of the transformation coefficients will give
rise to an infinite orthogonal matrix. The entries will be expressed as
Meixner polynomials: for parameters $\gamma>0$ and $0<c<1$ the Meixner
polynomial of degree $n$ is given by
\[
M_{n}\left(  x;\gamma,c\right)  =\,_{2}F_{1}\left(  -n,-x;\gamma
;1-c^{-1}\right)  ,
\]
with orthogonality relations
\begin{equation}
\sum_{x=0}^{\infty}\frac{\left(  \gamma\right)  _{x}}{x!}c^{x}M_{n}\left(
x;\gamma,c\right)  M_{m}\left(  x;\gamma,c\right)  =\frac{\delta_{mn}%
\,n!}{\left(  \gamma\right)  _{n}c^{n}\left(  1-c\right)  ^{\gamma}%
},\label{meixorth}%
\end{equation}
satisfying the second order difference equation
\begin{equation}
c\left(  x+\gamma\right)  f\left(  x+1\right)  -\left(  x+\left(
x+\gamma\right)  c\right)  f\left(  x\right)  +xf\left(  x-1\right)  =n\left(
c-1\right)  f\left(  x\right)  ,\label{meixdiff}%
\end{equation}
and the three-term recurrence
\begin{align}
xM_{n}\left(  x;\gamma,c\right)   &  =\frac{c\left(  \gamma+n\right)  }%
{c-1}M_{n+1}\left(  x;\gamma,c\right)  +\label{meix3}\\
&  \frac{n+\left(  n+\gamma\right)  c}{1-c}M_{n}\left(  x;\gamma,c\right)
+\frac{n}{c-1}M_{n-1}\left(  x;\gamma,c\right) \nonumber
\end{align}
(see the handbook of Koekoek and Swarttouw \cite{KS}, p.45).

Now consider the two orthogonal bases $\left\{  q_{0,n}:n\geq0\right\}  $ and
$\left\{  q_{1,m}:m\geq0\right\}  $ for $L^{2}\left(  \left(  0,\infty\right)
,x^{\alpha+1}dx\right)  $ where $q_{0,n}\left(  x\right)  =L_{n}^{\alpha
+1}\left(  \beta x\right)  e^{-\beta x/2}$ and $\,q_{1,m}\left(  x\right)
=L_{m}^{\alpha+1}\left(  \left(  2-\beta\right)  x\right)  e^{-\left(
2-\beta\right)  x/2}$ (for $0<\beta<2$, again not identical to previous usage
of this symbol). By use of the expansion (\ref{rescale}) for $\beta\neq1$ it
follows easily that
\begin{align*}
&  \Gamma\left(  \alpha+2\right)  ^{-1}\int_{0}^{\infty}q_{0,n}\left(
x\right)  q_{1,m}\left(  x\right)  x^{\alpha+1}\,dx\\
&  =\Gamma\left(  \alpha+2\right)  ^{-1}\int_{0}^{\infty}L_{n}^{\alpha
+1}\left(  \beta x\right)  L_{m}^{\alpha+1}\left(  \left(  2-\beta\right)
x\right)  \,e^{-x}x^{\alpha+1}\,dx\\
&  =\left(  1-\beta\right)  ^{n+m}\left(  -1\right)  ^{m}\frac{\left(
\alpha+2\right)  _{n}\left(  \alpha+2\right)  _{m}}{n!\,m!}\,_{2}F_{1}\left(
-n,-m;\alpha+2;\frac{\beta\left(  \beta-2\right)  }{\left(  1-\beta\right)
^{2}}\right)  ;
\end{align*}
and the $_{2}F_{1}$-sum equals $M_{n}\left(  m;\alpha+2,\left(  1-\beta
\right)  ^{2}\right)  $. Of course this calculation is only for illustration
and is not novel; the orthogonality of Meixner polynomials is a simple
consequence of that of the Laguerre polynomials and the values $\Gamma\left(
\alpha+2\right)  ^{-1}\int_{0}^{\infty}\left(  q_{0,n}\left(  x\right)
\right)  ^{2}x^{\alpha+1}\,dx=\beta^{-\alpha-2}\frac{\left(  \alpha+2\right)
_{n}}{n!}$ and the similar result for $q_{1,m}$.

\subsection{The transformation matrix}

We consider the analogous computation for the two bases \newline $\left\{
L_{m}^{\alpha+1}\left(  \frac{x}{\alpha+1}\right)  e^{-x/\left(  2\left(
\alpha+1\right)  \right)  }:m\geq0\right\}  $ and $\left\{  \phi_{n}%
:n\geq0\right\}  $ (from definition \ref{defphi}). The scaling for the former
is chosen to force equality at $m=n=0$.

\begin{proposition}
For $m,n\geq0$ let $A_{n}=\frac{1}{2}\left(  \frac{1}{\alpha+2n+1}+\frac
{1}{\alpha+1}\right)  ,u_{n}=\frac{n}{\alpha+n+1}$ and
\[
I\left(  n,m\right)  =\Gamma\left(  \alpha+2\right)  ^{-1}\int_{0}^{\infty
}L_{n}^{\alpha}\left(  \frac{x}{\alpha+2n+1}\right)  L_{m}^{\alpha+1}\left(
\frac{x}{\alpha+1}\right)  e^{-A_{n}x}\,x^{\alpha+1}\,dx,
\]
then $I\left(  0,0\right)  =\left(  \alpha+1\right)  ^{\alpha+2}$; $I\left(
0,n\right)  =0=I\left(  n,0\right)  $ for $n>0$; $A_{n}=\frac{1}{\left(
\alpha+1\right)  \left(  1+u_{n}\right)  }$ and
\begin{align*}
I\left(  n,m\right)   &  =\left(  \left(  \alpha+1\right)  \left(
1+u_{n}\right)  \right)  ^{\alpha+2}\frac{\left(  \alpha+3\right)
_{m-1}\left(  \alpha+2\right)  _{n-1}}{\left(  m-1\right)  !\,\left(
n-1\right)  !}\left(  -1\right)  ^{m-1}\left(  1-u_{n}^{2}\right) \\
&  \times u_{n}^{m+n-3}\,_{2}F_{1}\left(  1-n,1-m;\alpha+3;1-u_{n}%
^{-2}\right)  .
\end{align*}
\end{proposition}

\begin{proof}
In $I\left(  n,m\right)  $ change the integration variable to $s=A_{n}x$ and
let $\beta=\left(  A_{n}\left(  \alpha+2n+1\right)  \right)  ^{-1}%
=\frac{\alpha+1}{\alpha+n+1}$, then
\[
I\left(  n,m\right)  =\Gamma\left(  \alpha+2\right)  ^{-1}A_{n}^{-\alpha
-2}\int_{0}^{\infty}L_{n}^{\alpha}\left(  \beta s\right)  L_{m}^{\alpha
+1}\left(  \left(  2-\beta\right)  s\right)  e^{-s}s^{\alpha+1}ds.
\]
By formulae (\ref{rescale}) and (\ref{contig})
\begin{align*}
L_{n}^{\alpha}\left(  \beta s\right)   &  =\sum_{j=0}^{n}\frac{\left(
\alpha+2+j\right)  _{n-1-j}}{\left(  n-j\right)  !}\beta^{j}\left(
1-\beta\right)  ^{n-1-j}L_{j}^{\alpha+1}\left(  s\right) \\
&  \times\left(  \left(  1-\beta\right)  \left(  \alpha+n+1\right)
-(n-j)\right) \\
&  =\sum_{j=1}^{n}\frac{j\left(  \alpha+2+j\right)  _{n-1-j}}{\left(
n-j\right)  !}\beta^{j}\left(  1-\beta\right)  ^{n-1-j}L_{j}^{\alpha+1}\left(
s\right)  ,
\end{align*}
for $n>0$ while $L_{0}^{\alpha}\left(  \beta s\right)  =1$. By formula
(\ref{rescale})
\[
L_{m}^{\alpha+1}\left(  \left(  2-\beta\right)  s\right)  =\sum_{j=0}^{m}%
\frac{\left(  \alpha+2+j\right)  _{m-j}}{\left(  m-j\right)  !}\left(
2-\beta\right)  ^{j}\left(  \beta-1\right)  ^{m-j}L_{j}^{\alpha+1}\left(
s\right)  .
\]
Multiply the two sums and integrate with respect to $\Gamma\left(
\alpha+2\right)  ^{-1}e^{-s}s^{\alpha+1}ds$ and by use of the orthogonality
relations obtain
\begin{align*}
I\left(  n,m\right)   &  =A_{n}^{-\alpha-2}\left(  \alpha+2\right)
_{m}\left(  \alpha+2\right)  _{n-1}\left(  -1\right)  ^{m}\\
&  \times\sum_{j=1}^{\min\left(  m,n\right)  }\frac{j}{\left(  n-j\right)
!\left(  m-j\right)  !j!\left(  \alpha+2\right)  _{j}}\left(  \frac
{\beta\left(  \beta-2\right)  }{\left(  1-\beta\right)  ^{2}}\right)  ^{j},
\end{align*}
for $n>0$; and $I\left(  0,0\right)  =A_{0}^{-\alpha-2}=\left(  \alpha
+1\right)  ^{\alpha+2},I\left(  0,m\right)  =0$ for $m>0$. When $n>0$ let
$u_{n}=\frac{n}{\alpha+n+1}$ then $A_{n}=\left(  \left(  1+u_{n}\right)
\left(  \alpha+1\right)  \right)  ^{-1},\allowbreak\,\beta=1-u_{n}%
,\allowbreak2-\beta=1+u_{n}$. In the sum replace $j$ by $i+1$ (and $\left(
n-j\right)  !$ by $\left(  n-1\right)  !/\left(  1-n\right)  _{i}\left(
-1\right)  ^{i}$ and similarly for $\left(  m-1\right)  !$ ) to obtain the
claimed result.
\end{proof}

Certainly the integral could be computed for any scaling of the $\left\{
L_{m}^{\alpha+1}\right\}  $ basis, but our particular choice gives the neatest
formulae. Motivated by the value of $I\left(  n,m\right)  $ we write the following:

\begin{definition}
For $\alpha\geq0$ and $n=0,1,2,3\ldots$(and $u_{n}=\frac{n}{\alpha+n+1}$)
define the function
\[
h_{n}^{\alpha}\left(  x\right)  =u_{n+1}^{x}\left(  x+1\right)  M_{n}\left(
x;\alpha+3,u_{n+1}^{2}\right)  \text{ for }x\geq-1.
\]
\end{definition}

\begin{corollary}
For $n>0,$%
\begin{align*}
I\left(  n,m\right)   &  =\left(  \left(  \alpha+1\right)  \left(
1+u_{n}\right)  \right)  ^{\alpha+2}\frac{\left(  \alpha+3\right)
_{m-1}\left(  \alpha+2\right)  _{n-1}}{m!\,\left(  n-1\right)  !}\\
&  \times\left(  -1\right)  ^{m-1}\left(  1-u_{n}^{2}\right)  u_{n}%
^{n-2}h_{n-1}^{\alpha}\left(  m-1\right)  .
\end{align*}
\end{corollary}

Let
\begin{align*}
N_{m}  &  =\Gamma\left(  \alpha+2\right)  ^{-1}\int_{0}^{\infty}L_{m}%
^{\alpha+1}\left(  \frac{x}{\alpha+1}\right)  ^{2}e^{-x/\left(  \alpha
+1\right)  }\,x^{\alpha+1}\,dx\\
&  =\left(  \alpha+1\right)  ^{\alpha+2}\frac{\left(  \alpha+2\right)  _{m}%
}{m!},
\end{align*}
and recall from Theorem \ref{normphi} that $||\phi_{n}||^{2}=\left(
\alpha+2n+1\right)  ^{\alpha+3}\frac{\left(  \alpha+2\right)  _{n-1}}{n!}$
(dividing by the normalizing $\Gamma\left(  \alpha+2\right)  $). By the
orthogonality of the two bases the matrix $\left\{  \dfrac{I\left(
n,m\right)  }{||\phi_{n}||\,N_{m}^{1/2}}\right\}  _{n,m=0}^{\infty}$ is
orthogonal. The entries in row $0$ and column $0$ are zero except $I\left(
0,0\right)  =\left(  \alpha+1\right)  ^{\alpha+2}=||\phi_{0}||\,N_{0}^{1/2}$.
The row orthogonality demonstrates the orthogonality of the functions
$\left\{  h_{n}^{\alpha}\right\}  $; indeed $\sum_{m=0}^{\infty}N_{m}%
^{-1}I\left(  n,m\right)  I\left(  l,m\right)  =\delta_{nl}||\phi_{n}||^{2}$.
After some manipulation this implies for $n,l\geq1$ that
\begin{align*}
&  \sum_{m=1}^{\infty}\frac{\left(  \alpha+2\right)  _{m}}{m!}h_{n-1}^{\alpha
}\left(  m-1\right)  h_{l-1}^{\alpha}\left(  m-1\right) \\
&  =\delta_{nl}\frac{\left(  n-1\right)  !}{\left(  \alpha+2\right)  _{n-1}%
}\frac{\left(  \alpha+2\right)  ^{2}\left(  \alpha+n+1\right)  ^{2\alpha
+2n+4}}{\left(  \alpha+1\right)  ^{\alpha+4}\left(  \alpha+2n+1\right)
^{\alpha+3}n^{2n-3}}.
\end{align*}
For $n=l$ this can be verified from the Meixner polynomial relations: the sum
equals
\[
\left(  \alpha+2\right)  \sum_{x=0}^{\infty}\frac{\left(  \alpha+3\right)
_{x}}{x!}\left(  x+1\right)  u_{n}^{2x}M_{n-1}\left(  x;\alpha+3,u_{n}%
^{2}\right)  ^{2};
\]
then $\left(  x+1\right)  M_{n-1}\left(  x;\alpha+3,u_{n}^{2}\right)  $ can be
expanded in terms of \newline $\left\{  M_{j}\left(  x;\alpha+3,u_{n}%
^{2}\right)  :j=n-2,n-1,n\right\}  $ by use of formula (\ref{meix3}). The
coefficient of $M_{n-1}$ is $\frac{n\left(  \alpha+n+1\right)  }{\alpha+1}$
and together with (\ref{meixorth}) this produces the desired result. The
orthogonality of $\left\{  h_{n}^{\alpha}\right\}  _{n=0}^{\infty}$ can also
be established by exhibiting these functions as eigenfunctions of a
self-adjoint operator.

\subsection{A self-adjoint operator}

First let $M_{n-1}\left(  x;\alpha+3;u_{n}^{2}\right)  =u_{n}^{-x}g\left(
x\right)  $ and substitute in the second-order difference equation
(\ref{meixdiff}) to obtain:
\[
\left(  x+\alpha+3\right)  g\left(  x+1\right)  +\left(  \alpha+1\right)
g\left(  x\right)  +xg\left(  x-1\right)  =\left(  u_{n}+u_{n}^{-1}\right)
\left(  x+1\right)  g\left(  x\right)  ;
\]
replace $g\left(  x\right)  $ by $\frac{f\left(  x\right)  }{x+1}$ then $f$
satisfies
\[
\frac{x+\alpha+3}{x+2}f\left(  x+1\right)  +\frac{\alpha+1}{x+1}f\left(
x\right)  +f\left(  x-1\right)  =\left(  u_{n}+u_{n}^{-1}\right)  f\left(
x\right)  .
\]
It is easy to see that the operator
\[
\mathcal{D}f\left(  x\right)  =\frac{x+\alpha+3}{x+2}f\left(  x+1\right)
+\left(  \frac{\alpha+1}{x+1}-2\right)  f\left(  x\right)  +f\left(
x-1\right)
\]
defined for functions on $x=-1,0,1,2,\ldots$ such that $f\left(  -1\right)  =0
$ is symmetric for the inner product $\left\langle f_{1},f_{2}\right\rangle
=\sum\limits_{x=0}^{\infty}\dfrac{\left(  \alpha+2\right)  _{x+1}}{\left(
x+1\right)  !}f_{1}\left(  x\right)  f_{2}\left(  x\right)  $. Further
$\mathcal{D}h_{n-1}^{\alpha}=\dfrac{\left(  u_{n}-1\right)  ^{2}}{u_{n}%
}h_{n-1}^{\alpha} $ (for $n\geq1$), and $\dfrac{\left(  u_{n}-1\right)  ^{2}%
}{u_{n}}=\dfrac{\left(  \alpha+1\right)  ^{2}}{n\left(  \alpha+n+1\right)  },$
and so here is another proof of the orthogonality of the set $\left\{
h_{n}^{\alpha}:n\geq0\right\}  $.

Note that $\mathcal{D}$ can be written as
\begin{align*}
\mathcal{D}f\left(  x\right)   &  =\left(  f\left(  x+1\right)  -2f\left(
x\right)  +f\left(  x-1\right)  \right) \\
&  +\frac{\alpha+1}{x+2}\left(  f\left(  x+1\right)  -f\left(  x\right)
\right)  +\left(  \alpha+1\right)  \left(  \frac{1}{x+1}+\frac{1}{x+2}\right)
f\left(  x\right)  ,
\end{align*}
which is a discrete analogue of the radial Coulomb equation (\ref{radeq}). It
is clear that the difference equation
\begin{equation}
\mathcal{D}f\left(  x\right)  =-\lambda f\left(  x\right) \label{symmop}%
\end{equation}
with initial conditions $f\left(  0\right)  =1$ and $f\left(  -1\right)  =0$
has a unique solution for any $\lambda$. To find a formal expression for the
solution set
\[
f\left(  x\right)  =u^{x}\left(  x+1\right)  \sum_{j=0}^{\infty}a_{j}\left(
-x\right)  _{j}%
\]
with $u,a_{j}$ to be determined, and substitute in the equation (the series
terminates for each $x=0,1,2,\ldots$). After the usual manipulations we obtain
(for $j\geq0$):
\begin{gather*}
-u^{2}\left(  j+1\right)  \left(  j+\alpha+3\right)  a_{j+1}\\
+u\left(  u\left(  \alpha+2j+3\right)  +\alpha+1-\left(  2-\lambda\right)
\left(  j+1\right)  \right)  a_{j}\\
-\left(  \left(  u-1\right)  ^{2}+\lambda u\right)  a_{j-1}=0.
\end{gather*}
To make this a two-term recurrence, solve $u^{2}-\left(  2-\lambda\right)
u+1=0$ for $u$; when $\lambda<0$ select the solution with $0<u<1$.
Substituting $\lambda u=-\left(  u-1\right)  ^{2}$ in the recurrence results
in
\begin{equation}
-u^{2}\left(  j+1\right)  \left(  j+\alpha+3\right)  a_{j+1}+\left(
1+u\right)  \left\{  j\left(  u-1\right)  +u\left(  \alpha+2\right)
-1\right\}  a_{j}=0.\label{eqnaj}%
\end{equation}

First we handle the special case $\lambda=0,u=1$ with the solution
$a_{j+1}=\frac{2\left(  \alpha+1\right)  }{\left(  j+1\right)  \left(
j+\alpha+3\right)  }a_{j}$ and denote the corresponding solution $f$ of
(\ref{symmop}) by $h_{\infty}^{\alpha}$ where
\[
h_{\infty}^{\alpha}\left(  x\right)  =\left(  x+1\right)  \sum_{j=0}^{\infty
}\frac{\left(  -x\right)  _{j}}{\left(  \alpha+3\right)  _{j}}\frac{\left(
2\left(  \alpha+1\right)  \right)  ^{j}}{j!},
\]
a hypergeometric series of type $_{1}F_{1}$ (and resembling a Laguerre
polynomial). Second, for $u\neq1$ let $\gamma=\dfrac{1-u\left(  \alpha
+2\right)  }{u-1}$ then equation (\ref{eqnaj}) becomes
\[
-u^{2}\left(  j+1\right)  \left(  j+\alpha+3\right)  a_{j+1}+\left(
u^{2}-1\right)  \left(  j-\gamma\right)  a_{j}=0,
\]
with the corresponding solution of (\ref{symmop})
\[
f\left(  x\right)  =u^{x}\left(  x+1\right)  \,_{2}F_{1}\left(  -x,-\gamma
;\alpha+3;1-u^{-2}\right)  ,
\]
generally only meaningful for $x=-1,0,1,2,\ldots$. In this parametrization,
$u=\dfrac{1+\gamma}{\alpha+2+\gamma}$ and $\lambda=-\dfrac{\left(
\alpha+1\right)  ^{2}}{\left(  \gamma+1\right)  \left(  \alpha+\gamma
+2\right)  } $. (The terminating solutions of course were found already,
$\gamma=n$ for $h_{n}^{\alpha}$.) The asymptotic behavior of Meixner
polynomials and their zeros has been investigated by Jin and Wong
\cite{JW1},\cite{JW2}, a typical situation being the properties of
$M_{n}\left(  nx;\gamma,c\right)  $ as $n\rightarrow\infty$ for fixed
$x,\gamma,c$. But here $c$ depends on $n$, and a simple pointwise result is valid:

\begin{proposition}
$\lim\limits_{n\rightarrow\infty}h_{n}^{\alpha}\left(  x\right)  =h_{\infty
}^{\alpha}\left(  x\right)  $ for each $x\in\mathbb{R}$.
\end{proposition}

\begin{proof}
In $h_{n}^{\alpha}$ the argument of the \thinspace$_{2}F_{1}$ is
$1-u_{n+1}^{-2}=-\frac{\left(  \alpha+1\right)  \left(  \alpha+2n+3\right)
}{\left(  n+1\right)  ^{2}}.$ Then $u_{n+1}\rightarrow1$ and
\[
\frac{\left(  -x\right)  _{j}\left(  -n\right)  _{j}}{\left(  \alpha+3\right)
_{j}\,j!}\left(  -\frac{\left(  \alpha+1\right)  \left(  \alpha+2n+3\right)
}{\left(  n+1\right)  ^{2}}\right)  ^{j}\rightarrow\frac{\left(  -x\right)
_{j}}{\left(  \alpha+3\right)  _{j}\,j!}\left(  2\left(  \alpha+1\right)
\right)  ^{j},
\]
for each $j$ as $n\rightarrow\infty$. Since $\left|  \left(  -n\right)
_{j}/n^{j}\right|  \leq1$ the dominated convergence theorem applies to prove
the claim.
\end{proof}

For fixed $k=1,2,3,\ldots$ the first (in increasing order) $k$ zeros of
$h_{n}^{\alpha}$ tend to the first $k$ zeros of $h_{\infty}^{\alpha}.$

In this paper we gave an elementary proof of the strange orthogonality of
Laguerre functions associated with the wave functions of the Coulomb potential
and found an analogous orthogonality for Meixner polynomials.

\end{document}